\documentclass{article}
\def\be{\begin{equation}}
\def\ee{\end{equation}}
\def\bea{\begin{eqnarray}}
\def\eea{\end{eqnarray}}
\def\bk{{\mathbf k}}
\def\d{{\mbox{\rm d}}}

\begin{document}

%\begin{frontmatter}
\title{\bf Stable Bose-Einstein correlations}

\author{T. Cs\"org\H{o}$^1$, S. Hegyi$^1$ and W. A. Zajc$^2$\\[1ex]
    {\small $^1$ MTA KFKI RMKI, H - 1525 Budapest 114, P.O.Box 49, Hungary}\\
  {\small $^2$Dept. Physics, Columbia University,
        538 W 120th St, NY 10027 New York } 
}
%\author[KFKI]{T. Cs\"org\H o\thanksref{tamas}}
%\author[KFKI]{S. Hegyi\thanksref{hegyi}}
%\author[Columbia]{and W. A. Zajc\thanksref{bill}}
%\address[KFKI]{KFKI RMKI, H-1525 Budapest 114, POB 49, Hungary}
%\address[Columbia]{Dept. Physics, Columbia University, 
%	538 W 120th St, NY 10027 New York}

%\thanks[tamas]{Email: csorgo@sunserv.kfki.hu}
%\thanks[hegyi]{Email: hegyi@rmki.kfki.hu}
%\thanks[bill]{Email: zajc@nevis.columbia.edu}

%\date{\bf\Large Draft 0.6 }

\maketitle

\begin{abstract}
The shape of Bose-Einstein (or HBT) correlation functions
is determined for the case when particles are emitted
from a stable source, obtained 
after convolutions of large number of elementary random processes.
The two-particle correlation function is shown to have a 
{\it stretched exponential} 
shape, characterized by the L\'evy index of stability 
$ 0 < \alpha  \le 2$ and the scale parameter $R$. 
The normal, Gaussian shape corresponds to 
a particular case,  when $\alpha = 2$ is selected.
The asymmetry parameter of the stable source, $\beta$ is shown to be
proportional to the angle, measured by the normalized
three-particle cumulant correlations.
\end{abstract}

%\begin{keyword}
%correlations, elementary particle, heavy ion, statistical analysis, 
%L\'evy-stable distributions, two and three-particle correlations 
%\end{keyword}

%\end{frontmatter}
In high energy nuclear and particle physics, the space-time structure
of particle emitting sources is often investigated with the help
of the two-particle Bose-Einstein correlation functions. In heavy ion 
physics, these correlations are frequently called as HBT correlations
to honor the astronomers R. Hanbury Brown and R.Q. Twiss, who invented
a similar method~\cite{HBT2} in radio astronomy to measure 
the angular diameter of main sequence stars. 

The two-particle correlation  function $C_2(\bk_1,\bk_2)$ is defined as 
\begin{equation}
C_2(\bk_1,\bk_2) = \frac{N_2(\bk_1,\bk_2)}{N_1(\bk_1)\, N_1(\bk_2)}
\end{equation}
where $\bk_i$ is the momentum of particle $i=1,2$, and
$N_1(\bk_1)$ is the single particle invariant momentum distribution
(IMD), while $N_2(\bk_1,\bk_2)$ is the two-particle invariant
momentum distribution.

In this manuscript we highlight some of the results of ref.~\cite{cshz1},
where we have investigated in great detail 
the Bose-Einstein or HBT correlation functions under the following three

\underline{\it experimental conditions:}

{\it i) } 
{\it The correlation function tends to a constant
for large values of the relative momentum $q = k_1 - k_2$.}

{\it ii)}
{\it Near $|q|=0$, the correlation function deviates from its asymptotic, large $|q|$ value
in  a certain domain of  its argument. } 

{\it iii)} {\it The two-particle correlation function
 is related to a Fourier transformed
space-time distribution of the source.}
 
Condition {\it iii)} is satisfied if the propagation 
of identical boson pairs from a chaotic
(thermalized) source to the detector
can be described by a plane wave approximation; this is possible if Coulomb
and strong final state interactions as well as additional short range 
correlations e.g. caused by resonance decays can be 
corrected for or are negligible.

For clarity, let us consider first 
a one dimensional, factorized  toy model.
Let $x$ and $k$ stand for the coordinate and momentum variables,
respectively.  The model is defined by the emission function
\be
	S(x,k) =  f(x) \, g(k) ,
\ee
and the normalizations are
\be
	\int \d x \, f(x) \, = \, 1, \qquad\qquad
	\int \d k \, g(k)  =  \langle n \rangle, 
\ee 
where $\langle n\rangle$ stands for the mean multiplicity.
The single-particle spectrum is
\be
	N_1(k) = \int \d x \, S(x,k) = g(k).
\ee
If {\it iii)} is valid,
the Bose-Einstein symmetrized two-particle wave-function is 
\be
	\psi_{k_1,k_2}(x_1,x_2)  =  \frac{1}{\sqrt{2}}
	\left[\exp(i k_1 x_1 + i k_2 x_2) + 
		\exp(i k_1 x_2 + i k_2 x_1)\right].
\ee
In the Yano-Koonin formalism~\cite{yanokoonin}, 
the two-particle momentum distribution of chaotic sources 
is given as
\be
	N_2(k_1,k_2) = \int \d x_1 \d x_2 \, S(x_1,k_1) \, S(x_2,k_2) \,
		|\psi_{k_1,k_2}(x_1,x_2)|^2.
\ee
Let us introduce as auxiliary quantities the Fourier 
transformed source density distribution and  the relative momentum as
\be
	\tilde f(q)  =  \int \d x \, \exp(i q x) \,f(x),\qquad\quad
	q  =  k_1 - k_2. \label{e:fourier}
\ee
The two-particle Bose-Einstein correlation function is obtained as
\be
	C_2(k_1,k_2) = 1 + |\tilde f(q)|^2,
\ee
that measures the absolute value squared Fourier transformed
coordinate-space distribution function of the particle emitting source.

In physics, as well as in the theory of probability, 
the probability distribution of a sum of a large number
of random variables is one of the important problems,
and such distributions are frequently realized in Nature. 
Limit distributions characterize the probability distributions
of random processes in the limiting case when number of
the elementary independent random subprocesses 	tends to infinity.
In case of high energy nuclear and particle physics, for example,
the  position of emission of an observable particle is obtained as a sum of 
a large number of position shifts due to various parton-parton
scatterings, hadronization, rescattering of hadrons, and decay of
hadronic resonances:
\be
	x = \sum_n x_n
\ee
Hence the distribution of the sum $x$ is obtained
as an $n$-fold convolution,
\be
	f(x) = \int \d x_1 ... \d x_n f_1(x_1) ... f_n(x_n) 
	\delta(x - x_1 - x_2 .... - x_n)
\ee
Various forms of the Central Limit Theorem state, 
that under some conditions, the distribution of the sum of 
large number of random variables converges to a limit distribution.
In case of ``normal" elementary processes, the limit
distribution of their sum is the Gaussian distribution. 
This is one of the frequently encountered cases of limit
distributions.
	
	Stable distributions are precisely those limit distributions
that can occur in Generalized Central Limit theorems. Their study
was begun by the mathematician Paul L\'evy in the 1920's. 
A recent book by Zolotarev and Uchaikin~\cite{zol2} 
contains over 200 pages of applications of stable distributions 
in probabilistic models, correlated systems and fractals, 
anomalous diffusion and chaos,
physics, radiophysics, astrophysics, stochastic algorithms,
financial applications, biology and geology. Stable distributions
provide solutions to certain ordinary and fractional differential
equations. The breadth of their applications suggests  
that they can be considered as a class of special 
functions~\cite{zol2,zol1,nolan-summ}.

The Fourier transformed density distribution is usually called 
the {\it characteristic function} in mathematical statistics.
The stable distributions are frequently given in terms of
their characteristic functions.  
The reason for this is that the Fourier transform of 
a convolution is a product of the Fourier-transforms, 
\be
	\tilde f(q) = \prod_{i=1}^n \tilde f_i(q)
\ee
and limit distributions appear when the convolution
of one more elementary process does not change the 
shape of the limit distribution,  but it results only in
a modification of the parameters of the limit distribution. 
Hence the stable distributions have simple characteristic functions.
However, the explicit formulas 
describing the L\'evy stable source density distributions 
are known only in some special cases.
As of now, this is not an essential limitation as public domain
numerical packages exist that can be utilized to calculate
these source densities for any set of parameters~\cite{nolan-fitting}.

	Results of mathematical statistics yield a 
	 simple form for the characteristic function 
	of univariate and symmetric stable distributions, 
\be
	\tilde f(q)=\exp\left( i q \delta -|\gamma q|^\alpha\right), 
		\label{e:fqs}
\ee
	where the support of the density function $f(x)$ is $(-\infty,\infty)$.
	Deep mathematical results imply that the index of stability, $\alpha$, 
	satisfies the inequality $0 < \alpha \le 2$.  
	This parameter determines, for large modulus of the coordinates, the 
	L\'evy distributions. L\'evy laws with index of stability $\alpha$ 
	tend to power-laws and the exponent of the decay of 
	these distributions is given by
	$1+\alpha$, $f(x) \rightarrow |x|^{-1-\alpha}$ for $|x| \rightarrow \infty$.

	Although the proof that the shape given in eq.~(\ref{e:fqs})
	 is unique and that $0 < \alpha \le 2$	is rather complicated, it 
	is easy to show that these L\'evy distributions are indeed stable 
	under convolutions, 
\bea
  \tilde f_i(q) & = & \exp\left( i q \delta_i -|\gamma_i q|^\alpha\right),
  \qquad 
	\prod_{i=1}^n \tilde f_i(q) \, =  \, 
	\exp\left( i q \delta -|\gamma q|^\alpha\right) ,\\
	\gamma^\alpha & = &  \sum_{i=1}^n \gamma_i^\alpha, 
	\quad\qquad\qquad\qquad 
	\delta \, = \, \sum_{i=1}^n \delta_i, \label{e:gami}
\eea
	so after appropriate shifting and rescaling the stable distributions
	remain invariant.  Observe that eq. (\ref{e:gami}) 
	generalizes the well known quadratic addition rule 
	of variances of convoluted
	Gaussian distributions to the case of stable distributions.  

	In the following,  
	let us adopt the notation of Nolan~\cite{nolan-summ}. 
	Our choice corresponds to the $S(\alpha,\beta=0,\gamma = 
	R/2^{\frac{1}{\alpha}},\delta = x_0;1)$ convention. 
	In order to simplify the results, and to present results that
	are similar to the ones used in data fitting in high energy
	and nuclear physics, we have re-scaled the scale parameter
	$\gamma$ of the L\'evy distributions 
	and introduced a physical notation  as follows:
\be
 	R \, = \, 2^{\frac{1}{\alpha}}\,\gamma  ,
	\qquad\quad
	x_0  = \delta \qquad (\mbox{\rm if}\quad \beta = 0).\label{e:resc} 
\ee 
	In the chosen $S(\alpha,\beta,\gamma,\delta;1)$
	notational system, for symmetric $(\beta = 0)$ stable distributions, 
	the parameter $\delta$ coincides with $x_0$, 
	the location parameter of the distribution, characterizing the 
	position of particle production. (This parameter is irrelevant
	in Bose-Einstein correlation studies, we shall see that it cancels
	from both the two-  and the three-particle correlation functions.)
	In this notation, the one dimensional symmetric 
	L\'evy-stable distribution  yields the following,
	simple form of the two-particle BEC:
\be
	C(q) = 1 + \exp\left(-|q R|^\alpha\right).
		\label{e:c2stable}
\ee
	This  form that has an additional parameter, 
	the index of stability $\alpha$,
	as compared to the usual Gaussian (or exponential) distribution, 
	where the value of $\alpha$ is fixed to 2 (or 1).
	This result can be generalized straightforwardly
	to multidimensional expanding, core-halo type of systems.
	Two examples are given here, for more throughout discussion see 
	ref.~\cite{cshz1}.

	{\it Example a)}. 
	For collisions with non-relativistic energy, and a small
	duration of particle emission, symmetric L\'evy distributions
	yield the following Bose-Einstein correlation function~\cite{cshz1}:
\be
	C_2(k_1,k_2) = 1 + \lambda 
	\exp\left[ - (\sum_{i,j=1}^3 R_{ij}^2 q_i q_j)^{\alpha/2}\right].
\ee
	As usual, core-halo corrections~\cite{chalo} introduce 
	the intercept parameter $\lambda$. 
	The three-dimensional expansion of the core
	results in a multivariate decomposition of $q$. 
	All the fit parameters may depend on the mean momentum,
	$(\lambda,R_{ij}^2,\alpha) = 
		(\lambda({\bf K}),R_{ij}^2({\bf K}), \alpha({\bf K}))$,
	where ${\bf K} = 0.5 ({\bf k}_1 + {\bf k}_2)$.

	{\it Example b)}. 
	For collisions with very high energy, the particle emission
	process is a highly relativistic phenomena. In this case,
	the invariance of the emission function can be reflected
	if the longitudinally boost-invariant proper-time variable
	$\tau=\sqrt{t^2 - r_z^2}$ is utilized, and the space-time
	rapidity $\eta = 0.5 \log[(t + r_z)/(t-r_z) ] $ is also
	introduced as a hyperbolic, boost-additive coordinate.

	In a factorized form the Buda-Lund (BL) parameterization 
	assumes the following structure for the emission function~\cite{3d}:
\be
	S(x,k) = H_*(\tau) G_*(\eta) I_*(r_x,r_y),
\ee
	where the subscript $_*$ denotes an implicit momentum dependence.
	The effective proper-time and space-time rapidity distributions
	$H_*(\tau)$ and $G_*(\eta)$ are assumed to have a uni-variate 
	L\'evy distributions with indexes of stability
	$\alpha_=$ and $\alpha_{||}$, while $I_*(r_x,r_y)$ may have a bivariate
	L\'evy distribution with index $\alpha_\perp$. We use 
	the symbolic notation~\cite{cs-rev} of the invariant
	temporal and the parallel relative momentum differences,
	$Q_=$ and $Q_{||}$, being conjugated variables to 
	the space-time variables $(\tau,\eta)$.
	In these variables, the correlation function is 	
\bea
	C(k_1,k_2) &=& 1 + \lambda \exp\left(
		- 
		|R_= Q_=|^{\alpha_=}
		- 
		|R_\parallel  Q_\parallel|^{\alpha_\parallel}
		- 
		|R_\perp Q_\perp |^{\alpha_\perp}
		\right), \\
	Q_= & = & 
%		Q_0 \cosh(\overline{\eta}) - Q_z \sinh(\overline{\eta}) \, = \,
		 m_{t,1} \cosh(y_1 - \overline{\eta}) 
			- m_{t,2} \cosh(y_2 - \overline{\eta})\\
	Q_\parallel & = & 
%		Q_z \cosh(\overline{\eta}) - Q_0 \sinh(\overline{\eta}) 
		 m_{t,1} \sinh(y_1 - \overline{\eta}) 
			- m_{t,2} \sinh(y_2 - \overline{\eta})\\
	Q_\perp & = & \sqrt{Q_x^2 + Q_y^2}.
\eea
	In these equations, $m_{t,i} = 
	\sqrt{m^2 + {\bf k}_i^2}$ is the transverse
	mass, $y_i = 0.5\ln[(E_i + k_{z,i})/(E_i-k_{z,i})] $ is the rapidity 
	of particle $i$ and the fit parameter $\overline{\eta}$ 
	stands for the space-time
	rapidity of the point of maximum emissivity for particles with 
	a given fixed four-momentum $k_i$~\cite{cs-rev,cshz1}. 
	The three different indexes of stability satisfy the usual inequality
	$0 < \alpha_i \le 2$ for all $_i = ( \null_=, \null_{\parallel},
	\null_{\parallel})$.
	All the five fitted scale parameters, $\lambda$, $R_=$,
	$R_{\parallel}$, $R_{\perp}$	and $\overline{\eta}$,
	as well as the 
	three L\'evy indexes 
	$\alpha_=$, $\alpha_{\parallel}$ and $\alpha_\perp$
	may depend on the value of the mean momentum ${\bf K}$.

	Finally, let us  consider the case of three-particle Bose-Einstein 
	correlations.  If the particle emission is completely chaotic and
	the plane-wave approximation can be warranted, this reads as	
\bea
	C_3(1,2,3) \,  = \, 1 + |\tilde f(1,2)|^2 + |\tilde f(2,3)|^2
	 & +  &|\tilde f(3,1)|^2 +  \nonumber \\
 	 & +  & 	2 {\mathcal R} 	
			\tilde f(1,2) \tilde f(2,3) \tilde f(3,1).
\eea
	where the symbolic notation $\tilde f(i,j) \equiv \tilde f(k_i-k_j)
	\equiv \tilde f(q_{ij})$ has been introduced to simplify the equation.  
	The three-particle {\it cumulant} correlation function corresponds
	to the last term,
\be
	\kappa_3(1,2,3) = 2 {\mathcal R} \tilde f(1,2) \tilde f(2,3) 
		\tilde f(3,1),
\ee
	where the two-particle {\it cumulant} correlation function is
	defined as
\be
	\kappa_2(1,2) = |\tilde f(1,2)|^2 .
\ee
	Hence, the normalized and symmetrized ratio
\be
	w(1,2,3) = \frac{\kappa_3(1,2,3)}
		{2 \sqrt{\kappa_2(1,2) \kappa_2(2,3) \kappa_2(3,1)}}
\ee
	turns out to be a simple function of $\beta$, the asymmetry parameter:
\bea
	w(1,2,3) & =  &\cos \phi, \\
	\phi & = & \left\{\frac{\beta}{2} 
			R^\alpha \tan(\frac{\alpha \pi}{2}) 
			[\sum_{(i,j)}
			)
			|q_{ij}|^\alpha 
			\mbox{\rm sign}(q_{ij})]  \right\}
			\quad \mbox{\rm for}\quad  \alpha \ne 1, \label{e:cosfi}
\eea
	(for the special case of $\alpha=1$, see again ref.~\cite{nolan-summ}.)
	In the above equation, the summation is taken over the cyclic
	permutations, $(i,j) = (1,2)$, $(2,3)$, or $(3,1)$. 
	Note that the displacement parameter $\delta = x_0$ cancels
	from this result also. 

	In eq.~(\ref{e:cosfi}) all parameters are determined
	from the two-particle correlation function with the exception
	of $\beta$.  This parameter is limited to the range of
	$-1 \le \beta \le 1$, it is called 
	the asymmetry parameter of L\'evy distributions, and
	can thus be determined
	from the relative momentum dependence of the normalized three-particle
	cumulant correlation function $w$. Note that symmetric
	stable distributions correspond to the case of $\beta = 0$,
	hence in that case $\phi = 0$ and $w(1,2,3) = 1$ even at large
	relative momenta. 
	Thus all the essential parameters, $\alpha$, $\beta$
	and $\gamma  =R / 2^{\frac{1}{\alpha}} $ of stable source densities 
	can be reconstructed from two and three particle correlation data. 
	Let us emphasize that the result of eq.~(\ref{e:cosfi}) is
	valid only within the plane-wave approximation and neglecting
	possible partial coherence in the source~\cite{pcnhalo,cs-rev}. 

	{\it In summary}, we have determined the generic  structure 
	of Bose-Einstein correlations for the case when the coordinates
	of particle emission are obtained as convolution of many elementary
	subprocesses. Such processes can be attributed to parton-parton
	collisions, jet fragmentation, hadronization, rescattering
	and decay of hadronic resonances. 
	
	Our choice of multiple convolution of  various elementary probability laws  
	was motivated by a recent paper by A. Bialas~\cite{bialas}, that considered
	Bose-Einstein correlations for the case, when the radius of a Gaussian
	source fluctuates from event to event. Numerically, similar results
	were obtained by Utyuzh, Wilk and W{\l}odarczyk in ref.~\cite{wilk}
	when considering Bose-Einstein correlations for sources with 
	a fractal, power-law structure in space-time.

	We find that the general shape
	of the Bose-Einstein correlation functions is a {\it stretched
	exponential} form. This is an exact analytic result, valid
	for all values of the relative momentum $q$, if the particle
	production is described by a stable law. Thus a new, experimentally
	measurable parameter is introduced to HBT or Bose-Einstein
	correlation studies, the
	L\'evy index of stability, $\alpha$.  The corresponding 
	L\'evy stable distributions decay in the coordinate space 
	 as $|x|^{-1-\alpha}$ for large values of $|x|$, 
	see ref.~\cite{cshz1} for greater details.

	{\it Acknowledgments:} T. Cs. would like to thank the Organizers
	of the 2nd Warsaw meeting on correlations and resonances for creating 
	a meeting with  an excellent and inspiring 
	scientific atmosphere. This research was supported by an Alumni 
	Initiatives Award of the Fulbright Foundation, 
	an NWO - OTKA grant N25487, the Hungarian OTKA grants 
	(T024094, T026435), T034269, T038406, T043514,
	 the NATO PST.CLG.980086 grant and the MTA-OTKA-NSF grant INT0089462.

\vfill\eject
\end{document}